\documentclass{acm_proc_article-sp}
\include{graphix}

\newfont{\codefont}{pcrr at 9pt}
\newcommand\cplusplus{C\raisebox{.7ex}{$_{++}$} }

\title{A Scalable Stream-Oriented Framework for Cluster Applications}
\numberofauthors{1}
\author{
\alignauthor Tassos S. Argyros \hspace*{0.2in} David R. Cheriton\\
\affaddr{Distributed Systems Group}\\
\affaddr{Stanford University}\\
\email{\{Argyros, Cheriton\}@DSG.Stanford.EDU} }

\input{epsf}

\begin{document}

\maketitle

\begin{abstract}
This paper presents  a stream-oriented architecture for structuring
cluster applications. Clusters that run applications based on this
architecture can scale to tenths of thousands of nodes with
significantly less performance loss or reliability problems. Our
architecture exploits the stream nature of the data flow and reduces
 congestion through load balancing, hides latency behind data pushes
and transparently handles node failures. In our ongoing work, we are
developing an implementation for this architecture and we are able
to run simple data mining applications on a cluster simulator.
\end{abstract}

\section{Introduction}
One of the main characteristics of computing these days is the data
explosion; in fact, data nowadays grows at an exponential rate. What
is really surprising, however, is that the growth of the acquired
data surpasses the increase rate of the processing speed of today's
processors. As an example, the amount of data in GenBank (a genomics
database) doubles every 9 months, a much higher rate than the
18-month doubling period of processors (as dictated by Moore's law).
On the other hand, the processing of data is of imminent value,
since we continuously discover new ways to mine and process large
amounts of data to extract useful information. Fields such as
biology, physics, earth sciences and even marketing depend more and
more on mining huge amounts of data to break new ground and advance
forward.

Where all this leads to is that we need to find a way to process
ever-larger amounts of data with systems of reasonable cost. A
possible solution to this problem could be clusters of commodity
machines; after all, the cost of PC and network hardware keeps
falling at impressive rates. Thus, one would argue, all we have to
do to reach our goal is to buy more PCs, larger switches, and just
connect everything together.

Unfortunately, there are multiple problems when we try to follow
this approach. First, there are severe scaling problems. Indeed,
utilization of under 10\% of peak performance in large Linux
clusters has been reported \cite{}. The problem has only begin to
show up in current systems, since the majority of today's clusters
has relatively small sizes between 10 and 100 nodes, with only a
handful of systems being in the range over 1000 nodes. However, in
order to cope with the rapidly increasing amounts of data, we should
expect that clusters with many thousands of nodes will become the
rule rather than the exception in a few years; moreover, clusters
with 10,000 nodes or more should start appearing soon. One can only
imagine how severe scaling problems will appear in systems of such
size.

Apart from performance problems, today's cluster face a number of
other challenges. Node failures is one of them. Basic probability
theory dictates that the more nodes we have in a cluster, the
greater the chance that one of them will fail within a small period
of time. Identifying and correcting node failures is a hard task
that today usually requires human intervention. Moreover, writing an
application to run on a cluster, even a simple one, is much harder
than it seems. With today's technology one has to use some sort of
message passing interface or start making remote procedure calls to
transfer data. Such programming is difficult, time consuming and
hard to debug, especially in a parallel environment.

The question that naturally follows is what can we do about these
problems. More specifically, how should the data be send between the
nodes so that congestion and latency problems do not arise? How do
you deal with node and network failures that are inevitable in a
system of that size? What's the most effective way to interconnect
such a big network? And last but not least, how can one program and
debug such large scale applications? On a first thought, these
questions seem orthogonal to each other and inherent in every large
scale system.

However, we believe that this is not true. We claim that the problem
lies in the conventional way that today's systems are engineered,
and that an integrated solution to all of the above questions do
exist. Specifically, today's systems are build under the convention
that the applications determine data flow in an arbitrary way. What
this means is that a part of an application that runs on a cluster
node can potentially request data from any other node at any time;
and under this assumption, there is no systemic action that can be
taken to prevent scaling problems. Viewed from a point outside of
the application, data requests are random accesses, and random
access does not scale.

The solution that we propose is to build the whole system based on
\textit{streams}. More specifically, we propose to structure the
applications in a way that they read and write streams and build an
underlying framework that handles all the data flow issues. There
are numerous advantages that result from this approach. By infusing
the stream model into the applications, we can build a framework
that has enough knowledge of how the data flows within the cluster
to circumvent the scaling problems. Instead of using data requests
to send data from one node to the other (a latency-prone approach),
data can now be \textit{pushed} to its destination beforehand.

Moreover, this model significantly simplifies applications
programming. Issues that before needed to be handled by the
application, such as data flow and node failures, are now handled
transparently to the application by our framework. Since all data
exchange takes place using structures similar to files, there are no
complicated message-passing code that is hard to write, prone to
errors and sensitive to changes (e.g. in the size of the cluster).

Finally, node failures become manageable since the framework has
enough information to figure out what part of the data is lost and
proceed to corrective action. This way, every application that is
written using streams is robust to failures, at no expense to the
programmer.

 In the rest of this paper, we
present our on-going work which can be separated to two main tasks.
The first is to define the computational model of streams: their
programming representation and their semantics. The second task is
the implementation of the programming framework that includes
functions such as stream operations and node failures handling. We
begin by presenting the streams model.

\section{The Streams Model}
The goal of our system is to perform efficient processing on large
amounts of data. We refer to an independent and complete processing
application as a \textit{task}.

\begin{figure}[!ht]
\begin{center}
\epsfxsize=\columnwidth
\epsffile{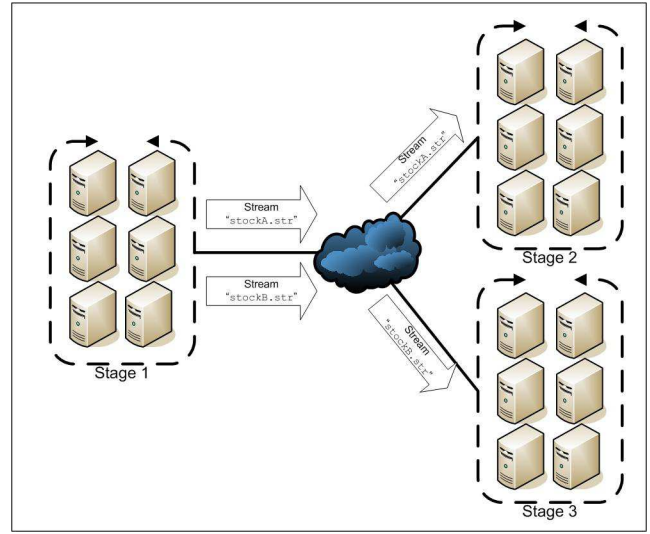}
\end{center}
\caption{In this example we have three stages. Stage 1 outputs
streams ``streamA.str" and ``streamB.str". Stage 2 inputs
``streamA.str" and stage 3 inputs ``streamB.str". Note that all
details of how stream data is delivered to nodes are handled by the
framework.}\label{fStages}
\end{figure}

An interesting observation is that in most cases a task can be
broken down to a number of independent operations that apply on data
as it flows through them. More specifically, we define a
computational \textit{stage} to be a logical processing unit: each
stage receives as input a number of streams, and it has the option
to apply some specific operations to the incoming streams. As an
example, a stage could specify that it needs to receive a particular
stream \textit{sorted} in some sense (we will define exactly what
this means in the next sections). The stage then can perform any
computation on the data units of a stream, and it can output the
results of the computation as another set of streams, for the next
stages to process. For an illustration of this concept see figure
\ref{fStages}

An important point here is to distinguish between the
\textit{declaration} and the \textit{definition} of a stage. More
specifically, a stage declaration is comprised by:
\begin{enumerate}
    \item the declaration of the input streams (which must be
    provided by previous stages),
    \item the declaration of the operations that should be applied on the input
    streams, and
    \item the declaration of the output streams.
\end{enumerate}
Additionally, a stage definition includes the actual function that
operates on the data units that a stream carries.

The reason why this distinction is important has to do with the fact
that the flow of data depends on the declaration of a stage solely.
In other words, we can determine with which other stages a stage
will exchange data. Thus, the layout of the data flow is now known
before the execution of the application, just by examining the
declarations of the stages.

A question that naturally arises is how are stages related to nodes?
This issue will be examined in the implementation section, but as a
simplification one can imagine that each stage is assigned to a set
of nodes with each node in the set running the \textit{same} piece
of code (specifically the stage function that operates on the stream
data units).

\subsection{Definition of Streams}
``Streams'' of data have been around since the first days of
computing. Usually, when one refers to a stream, she implies a data
flow with two main properties: sequentiality and uniformity. That
is, a stream can be abstracted as a sequence of \textit{data units}
with the restriction that these data units are of the same kind
(i.e. they share a common low-level data representation). As an
example, one could naturally define ``a stream of integers'' to mean
a data sequence with data units being 32-bit integers.

\begin{figure}[ht]
\begin{center}
\epsfxsize=\columnwidth \epsffile{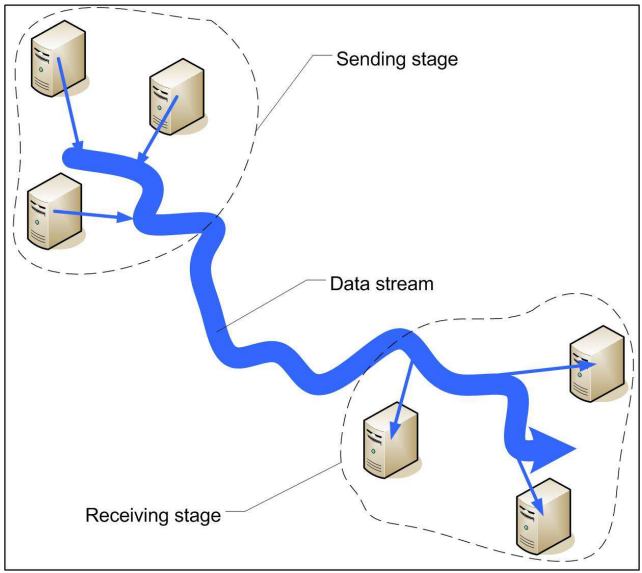}\end{center}
\caption{Three nodes send data to three other nodes using a stream
of data. Note that this is a high-level representation.
Implementation-wise, nodes use TCP pair-connections to transfer
data. }\label{fstream}
\end{figure}

In our model, we have expanded this classic definition of streams.
We raise the restriction of sequentiality, in the sense that in some
cases the exact order of data is of little importance. For example,
we may have three nodes sending data to another three nodes using a
single stream (see figure \ref{fstream}); because the three sending
nodes are not synchronized, order is difficult to be determined
here. The other extension to the streams definition is that our
model requires each data unit to be associated with a \textit{key}.
We can describe the key to be just a constant-length additional
field to each stream unit. Thus, each unit of a stream can be
abstractly described as a C struct:

\begin{center}
\codefont \fbox{\begin{minipage}{1.7in} \textbf{struct}
StreamDataUnit \{
\\\hspace*{0.2in} KeyType \textit{key};
\\\hspace*{0.2in} DataType \textit{data};
\\\};
\end{minipage}}
\end{center}

Although the \verb"data" field could change depending on the
DataType that a stream is associated to (i.e. the type
representation of a stream's data units), the type of the key,
\verb"KeyType" is the same for all streams. Intuitively, \verb"key"
is a numeric field that in some way acts as a representative for the
respective data in the various stream operations. In our current
implementation, KeyType is a fixed-size 64-bit integer.

Additionally, one soon realizes the need to uniquely identify every
stream. Towards this purpose, we define a global stream
\textit{namespace} where each stream is associated with a unique
name. We conventionally refer to streams with names that end in
``\verb".str"", but essentially any unique string is adequate.
Having all streams under a unique namespace is an important
simplification, since if a stage outputs a stream named
``\verb"sorted.str"", then another stage that is interested in
processing this stream just needs to ``ask" for the stream named
``\verb"sorted.str"" (more details on the programming interface are
provided in section \ref{sProgramming}).

\subsection{Stream Windows}\label{sWindows}
Streams have infinite length in theory, and usually unknown length
in practice. Since most applications simply need results before the
end of the stream, we need to find a way to start producing results
while a data stream still flows. As a solution to this, we propose
using stream \textit{windows} as computational units, upon which we
can perform operations and produce results. There are many ways to
define such a window; what we have used, and seems to make sense for
a large number of applications, is to define a window based on a
\textit{timestamp}. This timestamp is defined as a non-negative
integer and it is set by the programmer in the stage where the data
is produced. It has the additional requirement that it should be
increasing; that is, stream data units that are outputted first (at
a stage's output stream) are expected to have smaller timestamps
than the data units that follow. More formally, a window of
\textit{width} $T$ is defined to be all the data units of a stream
that have timestamps more or equal to $nT$ and less than $(n+1)T-1$,
for $n\in\aleph$. As an example, the first window will be all the
stream data units with timestamps $0$ to $T-1$ and the second will
have data units with timestamp from $T$ to $2T-1$.

Having defined the notion of window, we now proceed in defining the
stream operations.

\subsection{Operations on Streams}
As mentioned before, each stage can apply some operations on the
incoming streams. One could see these operations as a kind of
``queries" on streaming data. The reason why our framework provides
these operations (instead of having the applications programmer
providing them) has to do with the need of the framework being in
control of how the data flows. In a conventional approach, the
application programmer would need to program both the data flow
operations (e.g. using a message passing interface) and the
operations that apply on the data units. Our approach, in contrast,
is to distinguish these two programming tasks, provide the data flow
operations by the framework and leave up to the programmer the
simpler but more important task of programming the data operations.
In this way, we achieve both to give the framework full knowledge
and control of the data flow and to relieve the application
programmer from the hard task of programming how the data should
move from one node to the other.

In deciding which operations should be provided by the framework, we
wanted to select a relatively small operations set that would cover
the majority of our target applications. These operations act on
windows of data, as defined in the previous section. More
specifically, these operations and their semantics when applied to a
stream are:
\begin{description}
    \item[Sort] For each window, the data units of the operated stream will arrive at each node at
    a sorted order (either descending or ascending) based on the key
    value.
    \item[Group] For each window, the data of the operated stream will arrive at each node grouped by the key
    value, i.e. all stream data units of the same window that have the same key value will arrive
    together at each receiving node.
\end{description}

\begin{figure}[!ht]
\begin{center}
\epsfxsize=\columnwidth \epsffile{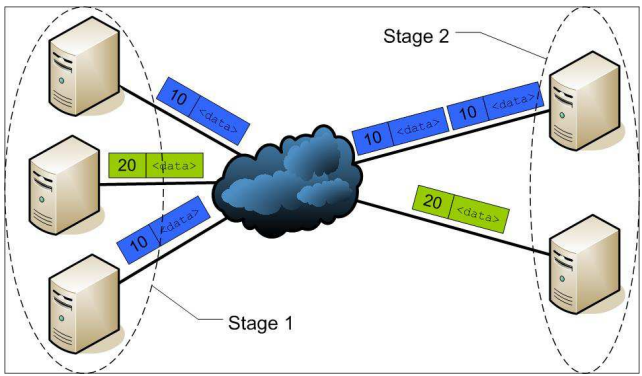}
\end{center}
\caption{An example of the Group operator in action. Here we have
two stages, with stage 1 sending data to stage 2 using a Grouped
stream. Note how all packets with key 10 end up in the same
node.}\label{fGroup}
\end{figure}

Figure \ref{fGroup} shows an example of the Group operation. There
we have packets of a stream flowing from the left three nodes to the
right three nodes. The stream has the Group operation applied to it,
and thus packets with the same key end up at the same node.

We argue that these two simple operators suffice for a surprising
large number of applications. Indeed, in many data processing
applications data is exchanged between the nodes of a cluster in
order to be grouped, to be sorted or to perform a seemingly
different operation that on a hindsight it is again based on sorting
or grouping.

There is a large number of operations that can be implemented using
our two basic operations combined with some in-stage computation.
Here are some examples:
\begin{itemize}
    \item \textbf{Join}. We can perform window-oriented joins between streams
    of data. By join we mean that we output a concatenation of data
    units from the joined streams if all these data units share the
    same key value. The way that this can be implemented is that we
    create a stage with input the streams that we would like to be
    joined with the Group operation applied. Then all
    the data units with the same key value will end up at the same
    node. All we have to do then is to see which of these key values
    span across all streams and output the respective data units.
    \item \textbf{Select}. A select operation can also be implemented.
    By select we mean that we want a stream to contain only key
    values that satisfy a specific criterion (perhaps a boolean
    formula). In order to implement this for a stream, we just add some code at
    the stage that outputs the stream that only allows a data unit
    to be outputted when it satisfies the mentioned criterion.
    \item \textbf{Aggregations}. Assume that we have a stream and we
    want to perform some aggregation operation on it, e.g. if the
    stream carries integers we may want to sum all integers in each
    window. One way to implement this is to define a stage that
    takes the stream to be aggregated as an input, defines another
    output stream that carries the aggregations and have that stream
    end up in the same node (by giving each aggregation the same
    key and applying the Group operator) where all the aggregations
    would sum up to produce the final number for each window.
\end{itemize}

One could argue whether these functions should be classified as
stream operations or simple applications. The point, however, is
that the Group and Sort operations with data units computations is a
much more powerful combination that what it seems.

\subsection{Load Balancing using Failure Management}
There are two reasons why a node may get overwhelmed with data:
\begin{itemize}
    \item The node may have lower performance compared to the other
    nodes due to a hardware/software problem. This is the classic
    definition of ``node failure".
    \item The actual amount of data send to the node may be much greater
    than the amount send to other nodes; thus although the node
    functions properly, it can not process all the incoming data as
    fast as the other nodes.
\end{itemize}

In our approach we do \textit{not} distinguish between these two
cases, although they may seem initially as two totally different
problems. We argue that by using the failure recovery mechanism for
both cases is the more appropriate tactic for the following reasons.

\begin{itemize}
    \item In many situations it is hard to distinguish between the
    two cases; thus by treating data overloads and failures
    differently, there is the risk of making a wrong decision and
    make the problem worse instead of fixing it.
    \item On an afterthought, the two problems are very similar in
    nature; indeed, the result of both is that a node is not able to
    finish a task that it has been assigned; thus the appropriate
    corrective action in both situations should be also very similar.
    \item Some load balancing mechanisms are anyway implemented into
    the failure management subsystem since after a failure occurs,
    the failure must be handled in a way that will not result in
    making other (non-failed) nodes overloaded.
    \item The design of the system is overall simpler and more
    effective when we can solve two important problems under a single
    mechanism.
    \item Under our approach, load balancing takes into
    consideration all possible causes of load unbalance, such as
    failures, bad data partitioning, non-uniformities in hardware or
    network and corrects them in run-time.
\end{itemize}

We thus believe that the above are significant advantages of our
approach compared to other conventional load balancing mechanisms.
Note that the definition of failure that we use, i.e. a node that
does not make adequate progress, means that our load balancing
mechanism can handle not only problems of data partitioning and
overloading, but also hardware and network problems (since e.g. a
bad network link would also cause slow node progress). More details
on the implementation of the above mechanisms are given in section
\ref{sFailure}.

\section{Programming Framework}\label{sProgramming}
In this section we present the programming framework that provides
the applications with the streams functionality. The approach that
we have taken is to implement the framework as a \cplusplus library,
that is compiled along with the application code.

More specifically, the programmer needs to perform two main tasks in
order to build a cluster application:
\begin{enumerate}
    \item The high level stage and stream declarations must be given. That is,
    the stages and the streams must be created and then each stage
    must specify which are the input and the output streams that it
    uses and what operations should be applied into the incoming
    streams.
    \item For each stage, the programmer must provide the actual function that performs the
    computation on the data units, that is reads the data from the
    incoming streams, modifies it and outputs the results to the
    outgoing streams.
\end{enumerate}

\subsection{Stage and streams declarations} Our aim is to declare
stages and streams in a single point within the code. In our current
implementation, the programmer needs to create a function where
stages and streams are declared as in the following example:

\begin{center}
\ttfamily \fbox{\begin{minipage}{\columnwidth} Ptr<Stage>
processingStage =
\\\hspace*{1.2in}stageManager->newStage( 10 );
\end{minipage}}
\end{center}

where the number $10$ declares the nodes to be assigned to this
stage. Also, a stream is declared similarly as

\begin{center}
\ttfamily \fbox{\begin{minipage}{\columnwidth}Ptr<Stream> dataStream
=
\\\hspace*{0.5in}streamManager->newStream( "data.str" );
\end{minipage}}
\end{center}

where the name of the stream is declared. After both a stage and a
stream is declared, the fact that the stage is associated with a
stream should be declared as well:

\begin{center}
\ttfamily \fbox{\begin{minipage}{\columnwidth}
    processingStage->newOutputStream( dataStream ); \\
    \normalfont or\\\ttfamily
    processingStage->newInputStream(dataStream,SORT);
\end{minipage}}
\end{center}

where in the second case, the second argument is the operation to be
applied to the input stream.

\subsection{Stage function definition}
 In our approach streams
share many characteristics with files. In order to read or write
data from a stream, one has to call the \verb"getStreamHandle()"
method that takes as an argument the string name of the stream and
whether it is an input or an output stream. Here are some examples:
\begin{center}
\ttfamily\small \fbox{\begin{minipage}{\columnwidth}
    StreamHandle distrStr = \\\hspace*{1in}getStreamHandle("distr.str",INPUT);\\
    StreamHandle mergeStr = \\\hspace*{1in}getStreamHandle("merge.str",OUTPUT);
\end{minipage}}
\end{center}

Using the \verb"StreamHandle" objects one can get or send packets
from/to the respective streams. For instance, the expression
\\\begin{center} \verb"data = getPacket( distrStr );"\\ \end{center}
would get the next packet from the queue of the stream
``\verb"distr.str"". Similarly, in order to send a packet to the
stream ``\verb"merge.str"" one simply has to issue the command
\begin{center}\verb"sendPacket(mergeStr, data);"\end{center}
In both these examples, data is of type \verb"DataUnit", which is
the class representation of a stream packet.

Note that in our implementation there is also a notification
mechanism for signaling the stage function when new packets arrive.

\section{Implementation}
We are currently implementing a framework that will support cluster
applications that use the stream model. To test and evaluate this
framework we are also developing a simulator of a cluster. In the
next sections we present the details of the framework implementation
and the simulation environment.
\subsection{Simulation Environment}
As mentioned before, we have developed a simulation environment to
test and develop our framework.

The lower layer of the simulator is the network, where nodes and
their interconnect are simulated and a basic mechanism to exchange
packets exists. Packet traffic is implemented using queues for each
node and event notifications.

We are currently progressing towards implementing the full
functionality of the framework as described in this paper. As a
future work, we plan to deploy it in a small-size real cluster for
further development and testing.

\subsection{Data flow}
As mentioned before, an important advantage of the stream model is
that data can be pushed to its destination, instead of having to be
requested in advance. In order to do this, we are using TCP
connections; actually, TCP fits our needs pretty well since it is
stream-oriented in nature.

There is the issue of how we create and manage the connections from
the application point of view. The case in most of today's systems
is that the programmer needs to explicitly identify which
connections must be created and between which nodes specifically
(e.g. one may need to specify the IP's of the nodes). However, in
our implementation after the declaration of stages and streams is
done, a connection is created automatically by the framework for
every pair of nodes that exchange data using a specific stream. The
ID's of the connecting nodes and all the other network details are
hidden behind the simple specifications of stages and streams.

To analyze this in more detail, when a stage is declared, the number
of nodes it is assigned to is also declared. Each of the assigned
nodes receive an initialization packet that includes the streams
that it inputs and outputs along with the IP's of the nodes that it
should be connected through these streams. Then the node
\textit{opens} a TCP connection with each node that is connected
through a input stream and \textit{listens} for a connection from
each node that is connected through an output stream.

When a node has made all the connections for its input and output
streams, it can start its computation. As soon as there is some
results from a stage, the nodes of that stage \textit{push} the data
to its destination; for flow management we use the standard TCP flow
control mechanisms. Consecutively, as soon as the next stages
receive the first data packets, they start processing data and
producing results.

\subsection{The Control Process}\label{sControl}
As soon as one tries to implement some control features, such as
failure management and load balance mechanisms, it becomes evident
that we need to synchronize the cluster nodes on some decisions. As
an example, in order to classify a specific node of a stage as
failed, \textit{all} nodes of the previous stage must agree on that
fact; otherwise we may end up in the unfortunate situation where
some part of the data still goes to the failed node (by the nodes
that do not see it as failed) while other nodes (that consider it
failed) send related data elsewhere.

In order to implement such features, a node is automatically chosen
to host a \textit{control process} and it is called the
\textit{control node}. This process communicates with the nodes of
each stage, receives data and feedback from them and makes decisions
such as whether a node should be considered failed or not (failure
management) or how the data should be partitioned to be send to the
receiving nodes (load balancing).

\begin{figure}[!ht]
\begin{center}
\epsfxsize=2.8in \epsffile{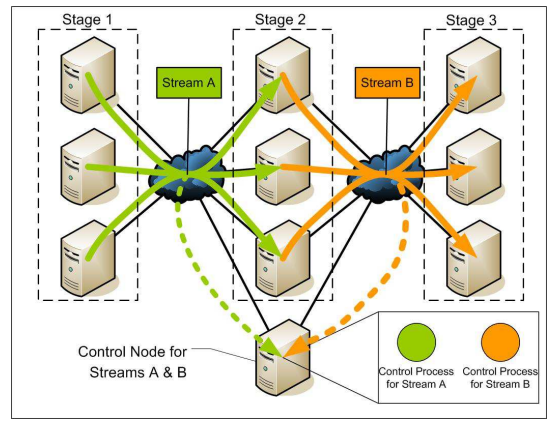}
\end{center}
\caption{Here we have three stages (1,2, and 3) and two streams (A
and B). The thin lines represent network connections, while thick
lines show the logical flow of the data. Note that we only have one
control node that runs two control processes for the two streams of
the cluster.}\label{fControl}
\end{figure}

Note that we can have one control node per stage or one control node
for all stages or something in between. In other words, a physical
node can run as many control processes as it can manage; we expect
that the control tasks will not be demanding in terms of processing
power and bandwidth requirements and thus a single node can act as a
control process for many stages. For an example of this fact, see
figure

\subsection{Windows-based Computation and Implicit State}
In section \ref{sWindows} we argued on the need to have windows as a
computation unit. Implementation-wise, what this means is that our
total computation task is actually a sequence of small window
computations. Under this model, state is directly related to
windows: the framework assumes that the data of a window are its
implicit state. Therefore, during a window computation the
\textit{only} data (state) that needs to be preserved is the data of
the \textit{current} window. This model enables us to perform load
balancing and failure management transparently, since the
application does not have to declare explicitly what state should be
recovered in case of a failure.

\subsection{GROUP Operation}
In the Group operation three entities are involved: the sending
stage, the receiving stage and the Group'ed stream. Remember that
the semantics of the Group operation is that we need all stream data
units with the same key to end up at the same node.

The main challenge in implementing the Group operation is how to
partition the keys in a way that no node gets either overloaded (by
receiving too much data) or stays idle (by receiving too little
data). Related work in this field proposes to make a pass over the
data before deciding on how to distribute it. However, in our case
this is too expensive and often impossible since we do not have the
luxury to store the (possible endless) incoming stream data in order
to determine the distribution; let alone the fact that determining
the actual distribution would require sorting the data, which brings
us back to our initial problem of how to split the data.

To tackle this problem we implement the following strategy. As soon
as some data is produced in the sending nodes, instead of streaming
it to the next stage we buffer it and measure its distribution. The
sending nodes use this information to determine how the data should
be split among the nodes. After we begin sending data, we consider
any overloaded nodes as \textit{failed} and use our failure handling
mechanism to redistribute the data of these nodes.

Let's see the above procedure with some more detail. First, we
partition the data units using a hash function. Note that we
partition based on the keys of the data units, and thus we ensure
that data units with identical keys end up in the same node. A
problem that arises is that we do not know the size of each
partition in advance; nor we can assume that all partitions have
equal sizes. The way we solve this issue is by making the number of
partitions a \textit{multiple} of the number of nodes of the
receiving stage; in other words, we hash the data to a number of
buckets that is many times the number of the nodes that these
buckets will end up to. By carefully assigning many buckets per node
we are able to circumvent the problem of buckets having different
sizes.

However, we can not make a proper assignment of buckets to nodes if
we do not know the size of each bucket. In order to get an estimate
of this, we do not start sending data as soon as it is available;
instead, we partition the data locally until a predetermined amount
of data has been partitioned\footnotemark{Explain how this is
determined.}. Then, the sending nodes communicate the partition
sizes to the respective stage control process (see section
\ref{sControl}). The control process executes an algorithm that
figures out the optimal split of the data (see Appendix \ref{algo})
and returns this information to the nodes that immediately start
sending the data to the next stage.

What we achieve with this process is a very good starting point in
the computation of the first window. Moreover, before we begin
computing the next window we repeat a similar process; but now we
can use the actual distribution of the data send in the previous
window to best determine how we should split the data in the next
window.

However, what if in the middle of a window computation the
distribution changes unexpectedly? For example, if we are dealing
with a stock trading processing system, a sudden increase in the
trading of a specific set of stocks that end up in the same node may
overload that node and hold back the whole computation. As
mentioned, our solution in this case is to handle overloads as
failures. We examine this mechanism in detail in section
\ref{sFailure}.

\subsection{SORT operation}\label{sort}
The Sort operation share many details with the Group operation. We
will describe here in which ways they are similar and we will focus
on the points that they differ.

\begin{figure*}[!t]
\begin{center}
\epsfxsize=5in \epsffile{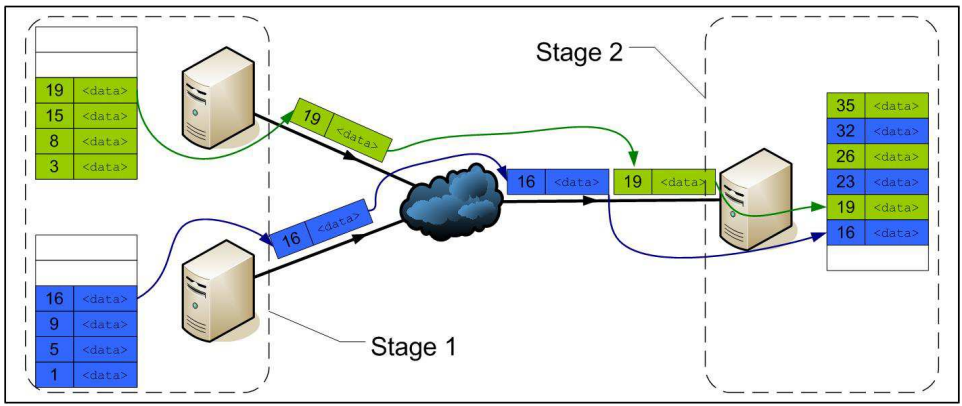}
\end{center}
\caption{Sorting by merging sorted streams. In order to receive a
data unit from the incoming stream, the node in stage 2 chooses the
packet with the highest key from all the incoming connections of the
specific stream.}\label{fSort}
\end{figure*}

What is traditionally used for sorting data using a cluster of
machines is to partition the data into key ranges and assign each
range of keys to a specific machine. We argue that under our stream
model we can sort in a much more efficient way. More specifically,
there are two ways to sort data using a distributed system:
\begin{enumerate}
    \item We must ensure that each node receives keys that belong to
    a particular range, e.g. [min\_value, max\_value). After each node
    receives all data with keys in that range, it sorts them. Then,
    whenever some other process needs sorted data that fall into
    that range, it queries that node. This is the conventional way
    of sorting data in a cluster.
    \item An alternative way is to have the sending nodes sort
    whatever piece of data they have and then \textit{stream out in sorted
    order} the data to the next stage. Each node of the next stage
    will simply need to \textit{merge} the incoming data; in other
    words, if the data is send in descending order, the nodes of the
    next stage simply need to read each time the data with the highest key
    from all the incoming connections. This concept is illustrated
    in figure \ref{fSort}.
\end{enumerate}

We argue that the alternative approach has significant advantages,
especially in a streaming environment like ours. As an example,
imagine an application that is only interested in the top 10\% of
the data. With the conventional approach, we need to send
\textit{all} data through the network since we do not know in
advance which part of the data will belong in the top 10\% (and
should be send out) and which not. In contrast, with our approach
the nodes will sort the data and they will only stream it out until
the next stage is no more interested in reading more data. In other
words, no more data is going to be send over the network than what
is essential for the computation.

\subsection{Failure Detection and Handling} \label{sFailure}
The first step in handling a failure is to detect it. However, since
we are building a distributed system, we must make sure that a node
must be considered either as failed or not failed by \textit{every}
other node that directly communicate with it. Note that since we
have partitioned our computation task into stages, the only subset
of nodes that are concerned with a failure are these of a previous
stage that send data through a stream to the failed node. Therefore
our model achieves to restrict the impact of a failure to only a
small subset of the total nodes of the cluster.

Our approach to detecting a failure is that each stage independently
decides on whether a node of a next stage is failed or not. The
nodes of a stage send periodic progress reports to the stage control
process for all nodes where they send data. If there is a node
failure then the progress report of all nodes will report that the
failed node does not seem to make enough progress; the control
process will then characterize this node as failed and it will
trigger the failure recovery mechanism.

The recovery mechanism is streams-oriented. This means that the
failure will be handled separately for each stream that sends data
to this node. The exact actions depend on whether this stream has a
Group or Sort operation assigned to it. We begin by looking how to
recover a stream with a Group operation.

Assume that we have the stream ``data.str" that is distributed
Grouped from stage $A$ to stage $B$. Node $n$ that receives data in
stage $B$ fails, and stage $A$ detects that. There will be a set of
key buckets that stage $A$ has associated with node $n$. In
implemented a failure recovery mechanism, the following issues need
to be solved:
\begin{enumerate}
    \item The key partition that was assigned to the failed node
    must be itself partitioned and redistributed to the other nodes
    without causing any other overloads.
    \item The data of the current window that was send to node $n$
    must have been stored somewhere (apart from node $n$).
    \item There must be a mechanism to detect which results $n$
    outputted before it failed, so that either stage B doesn't output duplicates or
    the stage that receives data from stage B can recognize and
    ignore duplicates.
\end{enumerate}

As soon as the control process decides to declare a node as failed,
it also makes a decision of how to repartition the key space. In
doing this, it assigns bigger part of the data of the failed node to
the those nodes that make the greatest progress. In this way, a node
failure actually results in a more even load across the nodes of
stage $B$. The algorithm to do that is similar with the one for the
initial partition (see Appendix \ref{algo}) and it is not presented
in detail in this paper.

The data is redundantly stored in the sending nodes. That is, as the
nodes stream out data they also buffer it as a back up in case of
failures. After the control process has declared a node failed, it
determines the new data partition and informs the nodes of stage $A$
that immediately start re-sending the buffered data following the
new partition.

We moreover need to detect which results were outputted by node $n$
before it failed, and which not. Since we can not count on the
failed node $n$ to give us this data, we should not assign this task
in the nodes of stage $B$ to determine this. Instead, with each data
unit that we output we include a \textit{sequence} number that
depends on the input data units that were used to produce the
specific result. The nodes of the stage after $B$ can check these
sequence numbers to simply ignore results that were send to them
before by node $n$.

\section{Related Work}
An other class of related work are database systems for streams.
Recently, several stream database systems have been build
\cite{STR03, aurora03, telegraph03}. However, most of them focus on
single-node environments and do not give any insight for
implementing such systems in a distributed environment. Moreover,
they are focused more on executing pre-determined queries on data
rather than acting as a framework for generic cluster applications.

The only database projets that is deals with distributed systems
issues in streams databases, to the extend of our knowledge, are the
Aurora* and Medusa projects \cite{aurora03, medusa03}. Note however
that these systems are distributed in the sense that two queries can
execute in two different nodes \--- there is no possibility of
executing a single query using multiple nodes, something that we
attempt to do in our work. Concerning the above systems, in
\cite{fault05}, fault recovery is examined in distributed stream
database systems. The paper follows a replication-based approach, in
contrast to our work. In \cite{avail05} several approaches on
failure recovery are presented and they are evaluated in a simulated
environment.

MapReduce \cite{Dean2004} is a programming model for writing cluster
applications using two functions: the Map function produces some
key/value pairs and the Reduce function merges the pairs that share
the same key. There are several points where this work differs than
ours. First, the streams programming model that we introduce is much
more general than simply using a Map/Reduce function pair, since it
also includes the notion of windows and operations. Moreover, our
system has load balancing mechanisms that are absent from the
MapReduce implementation. Finally, MapReduce writes intermediate
results to secondary storage (hard disks on nodes) in order to be
robust to failures. However, this approach has significant impact on
performance. Our aim, in contrast, is to handle failures on-the-fly
and use only memory to store intermediate data.

River \cite{river99} is a system that uses streams for load
balancing. However, load balancing with River can only take place at
specific points of the computation; namely, at points where each
data unit from one stage can be send to an arbitrarily chosen node
of the next stage under no restrictions (this excludes the Group and
Sort operations for instance). Our system implements a similar
mechanism for load balancing when there are no restrictions in the
flow of data; but we also achieve to have load balancing at
\textit{all} stages of our computation, a much more general result.
Also, like the River paper, our load balancing mechanism can balance
non-uniformities either in data distribution, in hardware or in the
network.

As part of the NOW Berkeley project \cite{now97}, there has been
some work on how can one use a cluster to sort efficiently
\cite{sort97}. The approach used is to split the data in ranges,
that (as we argued in section \ref{sort}) we believe that it is not
the best direction to perform a sort operation, at least in our
application model. Also, the authors make the explicit assumption
that the data follows a uniform distribution; in our work, we do not
make this assumption but rather we use data pre-processing to
approximate the actual distribution and load balance mechanisms to
cope with changes or bad approximations of the actual distribution.
Also, our algorithms are position to cope with the case that not all
source nodes carry the same amount of data, while this is an
essential assumption in the NOW-Sort paper.

There has also been some work on using streams in building faster
microprocessors. Specifically, the Imagine stream processor
\cite{imagine02} uses a stream model to bypass the memory bandwidth
bottleneck. Based on the Imagine stream processor, there is an
effort by the Merrimac project to build a full supercomputer that is
composed of stream processors \cite{merrimac03}. We believe that
this work differs in goals, assumptions and potential applications
from our work. More specifically, Merrimac aim is to achieve an
order of magnitude more TFLOPS than conventional supercomputers.
However, this greater computing capability requires pure scientific
applications that perform a large number of numerical computations
and access relatively small amounts of data; our framework, in
contrast, enables generic applications that process vast amounts of
data to achieve theoretical peak performance in commodity hardware.

\bibliographystyle{abbrv}
\bibliography{ref}

\appendix
\section{Sketch of Bucket Distribution Algorithm} \label{algo}
Consider the following problem: we have a set of buckets, $\{b_1,
b_2, \ldots, b_k\}$, each containing a number of items. Let these
numbers of items to be $n_1, n_2, \ldots, n_k$ respectively. We also
have $l$ nodes. We want to continuously group these buckets into the
$l$ nodes, meaning that each node will get a group of buckets in the
form $\{b_i, b_{i+1}, \ldots, b_{j-1}, b_j\}$. Our goal is to find
the bucket grouping that minimizes the variance of the number of
items that each nodes receives (in other words, it minimizes the
square of the distance of the number of items of each bucket from
the average).

In order to try all possible combinations, we would need exponential
time, since by applying some basic combinatorics we find that the
number of all possible instances of the problem is
$\binom{k+1}{l-1}$. However, there exists a dynamic algorithm
solution that solves the problem in polynomial time.

The basic idea of the algorithm is that if we have an optimal
distribution of $k$ buckets into $l$ nodes, and the $i$-th node
contains up to the $j$-th bucket, then the allocation of the first
$j$ buckets into the first $i$ nodes is also optimal. This is true
since if it was not optimal, the overall bucket allocation would not
be optimal as well.

Based on this observation, we construct a matrix $T$ of dimensions
$l \cdot k$. A matrix cell $T[i,j]$ contains the variance of the
optimal allocation for the first $j$ buckets into the first $i$
nodes. Also, we have calculated the average number of items per
node, $\mu$. From the mentioned optimality property, we can
calculate the $T[i,j]$ element using values only from the previous
column using the following formula:
\begin{equation}\label{minExp}
T[i,j] = \min_{k=1\ldots j-1}\left\{T[i-1, k] + (\sum_{t =
k+1}^{j}n_t - \mu)^2\right\}
\end{equation}

The variation of the optimal solution is $T[l,k]$. In order to find
the actual optimal bucket distribution, we construct a second matrix
$D$ of the same dimensions as $T$ and in each position we write the
minimum decision that we make from expression \ref{minExp}. Then we
can reconstruct the optimal distribution by going ``backwards" from
point $D[l,k]$ to $D[1,1]$.

The asymptotic running time of the algorithm is $O(lk)$ to for the
outer matrix loop and O(k) for calculating expression
\ref{minExp}\footnotemark{Instead of caclculating the sum in
\ref{minExp} each time, we can initially build an array and cache
the sums there. This way, the amortized time to calculate the sum is
$O(1)$.}. Thus the total running time is $O(lk^2)$. We have
implemented the algorithm and we have find out that it executes in
reasonable time for values of $l$, $k$ near 1000 (e.g. 1000 nodes
and 1000 buckets).

\end{document}